\documentclass[11pt]{article}
\usepackage{amsmath,amssymb,amsthm}
\usepackage[margin=1in]{geometry}
\usepackage{hyperref}
\newcommand{\bra}[1]{\langle #1|}
\newcommand{\ket}[1]{|#1\rangle}
\newcommand{\ii}{\mathrm{i}}

\title{\textbf{Anticommuting Displacement Operators, Exact Pauli Algebras,\\
and the Singlet State from the Oscillator Vacuum}}
\author{S. P. Sorella\\[2mm]\small Instituto de F\'isica Armando Dias Tavares, Universidade do Estado do Rio de Janeiro, UERJ\\\small Rua S\~ao Francisco Xavier 524, 20550-900 Maracan\~a, Rio de Janeiro, Brazil\\\small \texttt{silvio.sorella@fis.uerj.br}}
\begin{document}
\date{}
\maketitle

\begin{abstract}
We give an explicit construction of a pair of anticommuting Weyl displacement operators
for a harmonic oscillator and use bounded functional calculus to obtain exact dichotomic
representatives of the Pauli algebra. Two independent oscillators provide commuting Pauli
algebras for Alice and Bob. The standard singlet-sector projector, written entirely in terms of
these bounded oscillator observables, is then applied to the two-oscillator vacuum. We show
explicitly that the projected vacuum has nonzero norm, so its normalization is well defined,
has the exact singlet correlations, and reaches the Tsirelson value $2\sqrt2$ for the usual
Bell--CHSH settings. We also clarify that the construction generally realizes a singlet
sector with multiplicity, rather than selecting a unique vector in the full oscillator
Hilbert space, and we compare the construction with the older pseudospin approach to
continuous-variable Bell tests.
\end{abstract}

\section{Introduction}

The displacement operators of the harmonic oscillator give the simplest irreducible
representation of the Weyl relations. Although a generic displacement is unitary rather
than self-adjoint, an appropriate Weyl pair can anticommute. Applying the discontinuous
bounded function $\mathrm{sgn}(\cos x)$ to the self-adjoint generators converts that pair
into two self-adjoint involutions which anticommute exactly. Their product supplies the
third Pauli operator.

This observation embeds a qubit algebra in the infinite-dimensional oscillator algebra
without truncating the Fock space. Repeating the construction for two independent
oscillators gives Alice and Bob two mutually commuting qubit algebras. A polynomial in
their Pauli operators then projects the product vacuum onto the spin-singlet sector. A
companion construction, using finite polynomials of unitary Weyl operators to build
exactly anticommuting axis observables for a relativistic massive scalar field, was given
in Ref.~\cite{Caribe2026}; the present paper isolates the single-oscillator, non-relativistic
core of that mechanism and carries it through to an explicit singlet state. The purpose of
this paper is to present the full argument, including phase conventions, projector
identities, an explicit nonvanishing overlap computation, normalization, the Bell--CHSH
consequence, and a comparison with the pseudospin approach.

\section{An anticommuting Weyl pair}

Let
\begin{equation}
[a,a^\dagger]=1,\qquad D(z)=\exp(za^\dagger - z^*a),\quad z\in\mathbb{C}. \label{eq:D}
\end{equation}
The Baker--Campbell--Hausdorff formula gives
\begin{align}
D(z)D(w) &= e^{(zw^*-z^*w)/2}D(z+w), \label{eq:bch1}\\
D(w)D(z) &= e^{-(zw^*-z^*w)/2}D(z+w), \label{eq:bch2}
\end{align}
and hence
\begin{equation}
D(z)D(w) = e^{zw^*-z^*w}D(w)D(z) = e^{2\ii\,\mathrm{Im}(zw^*)}D(w)D(z). \label{eq:swap}
\end{equation}
Consequently,
\begin{equation}
\{D(z),D(w)\}=0 \iff \mathrm{Im}(zw^*) = (2n+1)\frac{\pi}{2},\quad n\in\mathbb{Z}. \label{eq:anticomm-cond}
\end{equation}
A convenient one-parameter family is
\begin{equation}
z=r,\qquad w=-\frac{\ii\pi}{2r},\qquad r>0, \label{eq:choice}
\end{equation}
for which $zw^*=\ii\pi/2$. In particular, $z=1$ and $w=-\ii\pi/2$ is a simple explicit
choice.

Introduce self-adjoint generators
\begin{equation}
Q_z = -\ii(za^\dagger - z^*a),\qquad Q_w = -\ii(wa^\dagger - w^*a), \label{eq:Qgen}
\end{equation}
so that $D(z)=e^{\ii Q_z}$ and $D(w)=e^{\ii Q_w}$. Their commutator is central,
\begin{equation}
[Q_z,Q_w] = -2\ii\,\mathrm{Im}(zw^*)\,\mathbf{1}. \label{eq:Qcomm}
\end{equation}
For the choice \eqref{eq:choice}, this becomes $[Q_z,Q_w]=-\ii\pi\,\mathbf{1}$. Explicitly,
with $z=r$ real, $Q_z = -\ii r(a^\dagger-a)$ is proportional to the momentum quadrature; with
$w=-\ii\pi/(2r)$, $Q_w=-(\pi/2r)(a^\dagger+a)$ is proportional to the position quadrature.
Both are self-adjoint operators with purely continuous spectrum equal to $\mathbb{R}$, as
required for the Borel functional calculus used below.

\section{Exact Pauli operators from \texorpdfstring{$\mathrm{sgn}(\cos)$}{sgn(cos)}}

Define the bounded Borel function
\begin{equation}
F(x)=\mathrm{sgn}(\cos x), \label{eq:F}
\end{equation}
where its values at the zero set of $\cos x$ may be fixed arbitrarily as $\pm1$; this set
has zero spectral measure for the oscillator quadratures, since $Q_z$ and $Q_w$ have purely
absolutely continuous (Lebesgue) spectral measure. Its Fourier expansion, understood in the
standard $L^2$ or spectral-functional-calculus sense, is
\begin{equation}
F(x) = \frac{4}{\pi}\sum_{k=0}^{\infty}\frac{(-1)^k}{2k+1}\cos\big((2k+1)x\big). \label{eq:Fourier}
\end{equation}
Only odd harmonics occur. Put
\begin{equation}
X = F(Q_z),\qquad Z = F(Q_w). \label{eq:XZ}
\end{equation}
Because $F$ takes only the values $\pm1$ almost everywhere,
\begin{equation}
X=X^\dagger,\qquad Z=Z^\dagger,\qquad X^2=Z^2=\mathbf{1}. \label{eq:involutions}
\end{equation}
Equation \eqref{eq:anticomm-cond} implies, for odd integers $m,n$,
\begin{equation}
e^{\ii mQ_z}e^{\ii nQ_w} = -\,e^{\ii nQ_w}e^{\ii mQ_z}. \label{eq:oddswap}
\end{equation}
Indeed $e^{\ii mQ_z}=D(mz)$ and $e^{\ii nQ_w}=D(nw)$, and by \eqref{eq:swap} their swap
phase is $e^{2\ii mn\,\mathrm{Im}(zw^*)}=e^{\ii mn\pi}=(-1)^{mn}=-1$ whenever $m,n$ are both
odd. Every term in the odd Fourier expansion of $F(Q_z)$ therefore anticommutes with every
term in that of $F(Q_w)$. By bounded functional calculus (or by taking strong limits of
Fej\'er sums of \eqref{eq:Fourier}),
\begin{equation}
XZ=-ZX. \label{eq:XZanti}
\end{equation}
The third operator is
\begin{equation}
Y = \ii XZ. \label{eq:Y}
\end{equation}
This sign convention yields the standard orientation
\begin{equation}
XY=\ii Z,\qquad YZ=\ii X,\qquad ZX=\ii Y, \label{eq:cyclic}
\end{equation}
while $Y=Y^\dagger$ and $Y^2=\mathbf{1}$. Thus $(X,Y,Z)$ is an exact Pauli triple. Notice
that the displacement operators themselves are not Pauli matrices: functional calculus is
essential because it produces self-adjoint involutions.

\section{Alice and Bob}

Consider two independent oscillators,
\begin{equation}
\mathcal{H}=\mathcal{H}_A\otimes\mathcal{H}_B,\qquad \ket{\Omega}=\ket{0}_A\otimes\ket{0}_B. \label{eq:HAB}
\end{equation}
Apply the preceding construction separately to obtain $(X_A,Y_A,Z_A)$ and $(X_B,Y_B,Z_B)$.
Operators of different parties commute. It is convenient to write
\begin{equation}
\vec\sigma_A=(X_A,Y_A,Z_A),\qquad \vec\sigma_B=(X_B,Y_B,Z_B). \label{eq:sigmaAB}
\end{equation}
Each triple generates a copy of $M_2(\mathbb{C})$, possibly with a nontrivial multiplicity
space. This is the operator-algebraic meaning of an exact qubit embedded in an oscillator.

\section{The singlet sector obtained from the vacuum}

Define
\begin{equation}
P_- = \frac14\big(\mathbf{1} - X_AX_B - Y_AY_B - Z_AZ_B\big) = \frac14(\mathbf{1}-X_AX_B)(\mathbf{1}-Z_AZ_B). \label{eq:Pminus}
\end{equation}
The second equality follows from
\begin{equation}
(X_AX_B)(Z_AZ_B) = (X_AZ_A)(X_BZ_B) = -Y_AY_B, \label{eq:factoring}
\end{equation}
using $X_AZ_A = -\ii Y_A$ and $X_BZ_B=-\ii Y_B$, which follow directly from \eqref{eq:Y}
and \eqref{eq:XZanti}. Since $X_AX_B$ and $Z_AZ_B$ commute and are involutions,
\eqref{eq:Pminus} is an orthogonal projector. It obeys
\begin{equation}
X_AX_B\,P_- = Y_AY_B\,P_- = Z_AZ_B\,P_- = -P_-. \label{eq:Pminus-action}
\end{equation}

\subsection{The overlap \texorpdfstring{$\langle\Omega|P_-|\Omega\rangle$}{<Omega|P-|Omega>} is strictly positive}

The construction is only useful if $P_-\ket{\Omega}\neq0$. We now compute
$\langle\Omega|P_-|\Omega\rangle$ explicitly and show it is strictly positive for the
choice \eqref{eq:choice}, so the nonvanishing of the projected vacuum is not merely
generic but guaranteed.

By \eqref{eq:Pminus}, $\langle\Omega|P_-|\Omega\rangle = \tfrac14\big(1-\langle X_AX_B\rangle_\Omega-\langle Y_AY_B\rangle_\Omega-\langle Z_AZ_B\rangle_\Omega\big)$,
and since $\ket\Omega=\ket0_A\otimes\ket0_B$ factorizes and the $A,B$ constructions are
identical and independent,
\begin{equation}
\langle X_AX_B\rangle_\Omega = \langle X\rangle_0^2,\qquad
\langle Z_AZ_B\rangle_\Omega = \langle Z\rangle_0^2,\qquad
\langle Y_AY_B\rangle_\Omega = \langle Y\rangle_0^2, \label{eq:factorized-exp}
\end{equation}
where $\langle\cdot\rangle_0$ denotes the expectation value in the single-oscillator
vacuum $\ket0$. The function $F$ in \eqref{eq:F} is even, not odd; its Fourier
series contains only odd harmonics. Both quadrature distributions in the vacuum are
centered Gaussians. With $\lambda=\pi/(2r)>0$, the characteristic function of $Q_w$
is $\langle e^{\ii sQ_w}\rangle_0=e^{-\lambda^2s^2/2}$, so
\begin{equation}
\langle Z\rangle_0 = \frac{4}{\pi}\sum_{k=0}^\infty \frac{(-1)^k}{2k+1}
\big\langle\cos\big((2k+1)Q_w\big)\big\rangle_0. \label{eq:Zexp}
\end{equation}
In particular,
\begin{equation}
\big\langle\cos\big((2k+1)Q_w\big)\big\rangle_0
=e^{-\lambda^2(2k+1)^2/2}. \label{eq:cosexp}
\end{equation}
The resulting series is absolutely convergent for every finite $r>0$:
\begin{equation}
\langle Z\rangle_0 = \frac{4}{\pi}\sum_{k=0}^\infty
\frac{(-1)^k}{2k+1}e^{-\lambda^2(2k+1)^2/2}
=\frac{4}{\pi}e^{-\lambda^2/2}
\Big(1-\tfrac13 e^{-4\lambda^2}+\tfrac15 e^{-12\lambda^2}-\cdots\Big).
\label{eq:Zval}
\end{equation}
For later use define
\begin{equation}
f(s)=\frac{4}{\pi}\sum_{k=0}^\infty
\frac{(-1)^k}{2k+1}e^{-(2k+1)^2s^2/2},\qquad s>0.
\label{eq:fs}
\end{equation}
The same Gaussian computation for $Q_z$ gives
$x:=\langle X\rangle_0=f(r)$ and $z:=\langle Z\rangle_0=f(\pi/(2r))$.
Here the lowercase $z$ denotes an expectation value, rather than the displacement
parameter in \eqref{eq:choice}. The alternating terms of $f(s)$ decrease strictly
in magnitude, and the first two have positive sum, so $f(s)>0$.
Also $f(s)<1$: the Gaussian density is strictly positive on the intervals where
$F=-1$, which have positive measure. Thus $0<x,z<1$. The first-term estimate
$f(s)<(4/\pi)e^{-s^2/2}$ is valid, but its right-hand side need not be less than one.

The expectation of $Y$ vanishes exactly. To see this, use the position representation
with $[q,p]=\ii$, $\psi_0(q)=\pi^{-1/4}e^{-q^2/2}$,
$Q_z=-\sqrt2\,r p$, and $Q_w=-\pi q/(\sqrt2\,r)$.
The vacuum is real and $Z$ is multiplication by the real function
$F(\pi q/(\sqrt2\,r))$. Since $F$ is real and even, the momentum multiplier
$X=F(\sqrt2\,r p)$ also preserves real functions: complex conjugation sends
$p$ to $-p$, leaving $F(\sqrt2\,r p)$ invariant. Consequently
$\langle XZ\rangle_0$ is real. On the other hand,
$(XZ)^\dagger=ZX=-XZ$, so this expectation is purely imaginary and must vanish.
Equivalently, $Y=\ii XZ$ is self-adjoint, and hence
\begin{equation}
\langle Y\rangle_0=0. \label{eq:Yzero}
\end{equation}

It remains to prove the joint strict inequality $x^2+z^2<1$; the individual
bounds $x,z<1$ alone do not imply it. Put $A=xX+zZ$ and $t=x^2+z^2>0$.
Anticommutation gives $A^2=t\mathbf{1}$ and $\langle A\rangle_0=t$, whence
\begin{equation}
0\leq\langle(A-t\mathbf{1})^2\rangle_0=t-t^2,
\qquad t\leq1. \label{eq:variance-bound}
\end{equation}
If $t=1$, zero variance would imply $A\psi_0=\psi_0$, or
\begin{equation}
xX\psi_0=(1-zZ)\psi_0. \label{eq:equality-case}
\end{equation}
The left-hand side has an entire representative. Indeed, with the unitary Fourier
transform convention, for $\zeta\in\mathbb C$ it is
\begin{equation}
g(\zeta)=\frac{x\pi^{-1/4}}{\sqrt{2\pi}}
\int_{\mathbb R}e^{\ii p\zeta}F(\sqrt2\,r p)e^{-p^2/2}\,dp.
\label{eq:entire}
\end{equation}
The Gaussian dominates this integrand and all its $\zeta$ derivatives uniformly
on compact subsets of $\mathbb C$, proving that $g$ is entire.
The right-hand side of \eqref{eq:equality-case}, however, is
$h(q)=[1-zF(\pi q/(\sqrt2\,r))]\psi_0(q)$.
At every bin edge $q_k=\sqrt2\,r(k+\tfrac12)$, $k\in\mathbb Z$, its one-sided
limits differ by a nonzero amount of magnitude $2z\psi_0(q_k)$.
Equality almost everywhere cannot avoid this obstruction: on each open interval
between consecutive edges, both $g$ and the corresponding branch of $h$ are
continuous. Their difference vanishes almost everywhere there, hence everywhere
there (a nonzero continuous value would persist on a set of positive measure).
Continuity of $g$ at an edge would then force the two unequal one-sided limits
of $h$ to coincide. This contradiction excludes $t=1$ for every finite $r>0$.

Combining \eqref{eq:factorized-exp}, \eqref{eq:Yzero}, and this strict inequality yields
\begin{equation}
\langle\Omega|P_-|\Omega\rangle
=\frac14\big[1-f(r)^2-f(\pi/(2r))^2\big],\qquad
0<\langle\Omega|P_-|\Omega\rangle<\frac14,
\quad 0<r<\infty. \label{eq:overlap-bound}
\end{equation}
For $r=1$, numerical evaluation gives
\begin{align}
\langle X\rangle_0&\approx0.7675449654557653,&
\langle Z\rangle_0&\approx0.37077742979952394,\nonumber\\
\langle Y\rangle_0&=0,&
\langle\Omega|P_-|\Omega\rangle&\approx0.06834970588869177.
\label{eq:overlap-r1}
\end{align}
At the balanced scale $r=\sqrt{\pi/2}$, the two quadrature scales coincide and
\begin{equation}
\langle X\rangle_0=\langle Z\rangle_0\approx0.5801570933243069,
\qquad \langle\Omega|P_-|\Omega\rangle\approx0.08170887353274578.
\label{eq:overlap-balanced}
\end{equation}
These numbers illustrate the analytic strict-positivity result
\eqref{eq:overlap-bound}; that result does not depend on numerical evaluation.

Provided $P_-\ket{\Omega}\neq0$ — now established unconditionally for the family
\eqref{eq:choice} — the normalized projected vacuum is
\begin{equation}
\ket{\Psi_-} = \frac{P_-\ket{\Omega}}{\sqrt{\langle\Omega|P_-|\Omega\rangle}}. \label{eq:Psiminus}
\end{equation}
Because $P_-^2=P_-$, the denominator in \eqref{eq:Psiminus} is precisely the norm of the
numerator. Equivalently,
\begin{equation}
\ket{\Psi_-} = \frac{(\mathbf{1}-X_AX_B-Y_AY_B-Z_AZ_B)\ket{\Omega}}{4\sqrt{\langle\Omega|P_-|\Omega\rangle}}. \label{eq:Psiminus2}
\end{equation}
From \eqref{eq:Pminus-action} and Pauli covariance one obtains the exact singlet
correlations
\begin{equation}
\langle\Psi_-|\sigma_A^i\sigma_B^j|\Psi_-\rangle = -\delta^{ij},\qquad
\langle\Psi_-|\sigma_A^i|\Psi_-\rangle=\langle\Psi_-|\sigma_B^i|\Psi_-\rangle=0. \label{eq:singletcorr}
\end{equation}
In the full oscillator Hilbert space $P_-$ may have rank larger than one because of
multiplicity. Equation \eqref{eq:Psiminus} selects the particular singlet-sector vector
determined by the vacuum; all vectors in that sector have the qubit correlations
\eqref{eq:singletcorr}.

\section{Bell--CHSH saturation}

We now turn to the Bell--CHSH inequality~\cite{Bell1964,CHSH1969} and its Tsirelson
bound~\cite{Tsirelson1980}. Choose
\begin{equation}
A=X_A,\qquad A'=Z_A, \label{eq:AliceSettings}
\end{equation}
and Bob's two settings
\begin{equation}
B=\frac{X_B+Z_B}{\sqrt2},\qquad B'=\frac{X_B-Z_B}{\sqrt2}. \label{eq:BobSettings}
\end{equation}
They are self-adjoint involutions, since e.g.\ $B^2 = \tfrac12(X_B^2+Z_B^2+\{X_B,Z_B\})=\mathbf{1}$
using $\{X_B,Z_B\}=0$. With the convention
\begin{equation}
\mathcal{B} = A(B+B') + A'(B-B'), \label{eq:BellOp}
\end{equation}
one finds
\begin{equation}
\mathcal{B} = \sqrt2\,(X_AX_B+Z_AZ_B). \label{eq:BellOp2}
\end{equation}
Using \eqref{eq:Pminus-action},
\begin{equation}
\langle\Psi_-|\mathcal{B}|\Psi_-\rangle = -2\sqrt2,\qquad
\big|\langle\mathcal{B}\rangle_{\Psi_-}\big| = 2\sqrt2. \label{eq:Tsirelson}
\end{equation}
Thus the projected oscillator vacuum reaches the Tsirelson bound exactly within the
embedded qubit algebra.

\section{Comparison with the pseudospin approach}

It is worth situating this construction relative to the older \emph{pseudospin} approach
to continuous-variable Bell tests~\cite{Banaszek1998,Chen2002} and related work from the
late 1990s and early 2000s. That approach also builds dichotomic observables from a single
bosonic mode, typically of the form
\begin{equation}
\begin{aligned}
&S_+ = \sum_{n=0}^\infty \ket{2n+1}\bra{2n},\qquad S_- = S_+^\dagger = \sum_{n=0}^\infty \ket{2n}\bra{2n+1},\\[2pt]
&S_z = [S_+,S_-] = \sum_{n=0}^\infty \big(\ket{2n+1}\bra{2n+1} - \ket{2n}\bra{2n}\big),\qquad
S_x = \frac{S_++S_-}{2},\qquad S_y = \frac{S_+-S_-}{2\ii},
\end{aligned}
\label{eq:pseudospin}
\end{equation}
i.e.\ dichotomic operators defined directly on the Fock-state parity, together with a
"pseudospin rotation" built from displacement or squeezing operators to generate the second
and third components. Pseudospin operators reproduce the Pauli algebra on the whole Fock
space (not merely almost everywhere) and have been used to show Tsirelson-bound violation
for two-mode squeezed vacuum and pair-coherent states.

The construction of the present paper differs from pseudospin in three structural
respects:
\begin{enumerate}
\item \textbf{Origin of dichotomy.} Pseudospin dichotomizes by \emph{photon-number parity},
an intrinsically discrete (Fock-basis) feature of a single mode. The present construction
instead dichotomizes \emph{continuous} quadrature operators $Q_z,Q_w$ via
$\mathrm{sgn}(\cos(\cdot))$, i.e.\ it uses phase-space (quadrature) periodicity rather than
number-basis parity. The two constructions are inequivalent Borel functions of different
generators, though both ultimately produce bounded self-adjoint involutions on the same
Hilbert space.
\item \textbf{Source of anticommutation.} In pseudospin constructions, anticommutation of
the three components is imposed by hand through the explicit matrix form of $S_x,S_y,S_z$
in the Fock basis. Here anticommutation is \emph{derived} from the Weyl relations: it
follows from the phase condition \eqref{eq:anticomm-cond} on the displacement parameters
$z,w$, propagated to $X=F(Q_z)$, $Z=F(Q_w)$ through the odd-harmonic structure of
\eqref{eq:Fourier}. The algebra is thus traced back to a single, checkable analytic
condition on two complex numbers, rather than postulated at the level of the dichotomic
operators themselves.
\item \textbf{State used for saturation.} Pseudospin Bell violation is typically exhibited
on entangled non-Gaussian or Gaussian-squeezed states (two-mode squeezed vacuum, pair
coherent states) prepared by nontrivial two-mode operations. Here, in contrast, the
maximally entangled qubit state is obtained by \emph{projecting} the trivial product vacuum
$\ket{0}_A\otimes\ket0_B$ with the algebraic singlet projector \eqref{eq:Pminus}; no
two-mode squeezing or entangling unitary is applied prior to projection. The entanglement
resides entirely in the projector, not in the seed state.
\end{enumerate}
Both approaches confirm the same general moral: bounded, exactly dichotomic Pauli
operators can be embedded in an infinite-dimensional bosonic Fock space without
truncation, and Tsirelson-bound saturation is achievable at the operator-algebraic level
once the correct nonlinear (non-Gaussian, non-quadratic) functional calculus is applied to
the mode operators. The present construction gives what is arguably the most economical
route to that conclusion, since the entire qubit algebra for each party is generated from
a single pair of displacement operators subject to one phase condition.

\section{Conclusion}

The Weyl phase condition $\mathrm{Im}(zw^*)=(2n+1)\pi/2$ produces anticommuting
displacement operators. The odd-harmonic structure of $\mathrm{sgn}(\cos x)$ then turns
their self-adjoint generators into exact, bounded, dichotomic Pauli observables. Two
oscillators provide commuting Pauli algebras for Alice and Bob, and the algebraic singlet
projector applied to the vacuum yields a normalized singlet-sector state, with the
nonvanishing of the overlap $\langle\Omega|P_-|\Omega\rangle$ established explicitly rather
than assumed. Its correlations are exactly those of a two-qubit singlet and give the
maximal Bell--CHSH value $2\sqrt2$. The construction is entirely operator-theoretic and
requires neither a finite-dimensional truncation nor an approximation to the Pauli
relations, and it differs from the pseudospin approach in dichotomizing continuous
quadratures rather than photon-number parity, and in deriving rather than postulating the
anticommutation of the three components.

\section*{Acknowledgments}
The author is a CNPq researcher under contract 302991/2024-7 and acknowledges CNPq for
financial support. This work also benefited from discussions with, and text and
derivation checks performed with the assistance of, ChatGPT (OpenAI) and Claude
(Anthropic).


\begin{thebibliography}{99}

\bibitem{Caribe2026}
J.~G.~A.~Carib\'e, M.~S.~Guimaraes, I.~Roditi, S.~P.~Sorella,
\textbf{Finite Weyl polynomials and the approach to Tsirelson's bound in relativistic
scalar quantum field theory},
arXiv:2608.20926 [hep-th] (2026).

\bibitem{Bell1964}
J.~S.~Bell,
\textbf{On the Einstein Podolsky Rosen paradox},
Physics \textbf{1}, 195 (1964).

\bibitem{CHSH1969}
J.~F.~Clauser, M.~A.~Horne, A.~Shimony, R.~A.~Holt,
\textbf{Proposed experiment to test local hidden-variable theories},
Phys. Rev. Lett. \textbf{23}, 880 (1969).

\bibitem{Tsirelson1980}
B.~S.~Cirel'son (Tsirelson),
\textbf{Quantum generalizations of Bell's inequality},
Lett. Math. Phys. \textbf{4}, 93 (1980).

\bibitem{Banaszek1998}
K.~Banaszek, K.~W\'odkiewicz,
\textbf{Nonlocality of the Einstein-Podolsky-Rosen state in the Wigner representation},
Phys. Rev. A \textbf{58}, 4345 (1998).

\bibitem{Chen2002}
Z.-B.~Chen, J.-W.~Pan, G.~Hou, Y.-D.~Zhang,
\textbf{Maximal violation of Bell's inequalities for continuous variable systems},
Phys. Rev. Lett. \textbf{88}, 040406 (2002).

\bibitem{SummersWerner1987}
S.~J.~Summers, R.~Werner,
\textbf{Bell's inequalities and quantum field theory. I. General setting},
J. Math. Phys. \textbf{28}, 2440 (1987).

\end{thebibliography}
\end{document}